\documentclass[10pt]{article}
\usepackage{ametsoc2col}
\newcommand{\myabstract}{
  Variations in zonal surface temperature gradients and zonally asymmetric 
  tropical overturning circulations (Walker circulations) are examined 
  over a wide range of climates simulated with an idealized atmospheric 
  general circulation model (GCM). The asymmetry in the tropical climate is 
  generated by an imposed ocean energy flux, which does not vary with climate.
  The range of climates is simulated by modifying the optical thickness of 
  an idealized longwave absorber (representing greenhouse gases). \\

  The zonal surface temperature gradient in low latitudes generally decreases 
  as the climate warms in the idealized GCM simulations.
  A scaling relationship based on a two-term balance in
  the surface energy budget accounts for the changes in the zonally asymmetric
  component of the GCM-simulated surface temperature gradients. \\

  The Walker circulation weakens as the climate warms in the 
  idealized simulations, as it does in comprehensive simulations of climate 
  change.
  The wide range of climates allows a systematic test of energetic arguments 
  that have been proposed to account for these
  changes in the tropical circulation. The analysis shows that a scaling
  estimate based on changes in the hydrological cycle (precipitation rate
  and saturation specific humidity) accounts for the simulated changes in the 
  Walker circulation. However, it must be evaluated locally, 
  with local precipitation rates. If global-mean quantities are used,
  the scaling estimate does not generally account for changes in the 
  Walker circulation, and the extent 
  to which it does is the result of compensating errors
  in changes in precipitation and 
  saturation specific humidity that enter the scaling estimate.
}

\newcommand  {\about} {\mathop{\sim}\!}

\begin{document}
%
%%%%%%%%%%%%%%%%%%%%%%%%%%%%%%%%%%%%%%%%%%%%%%%%%%%%%%%%%%%%%%%%%%%%%
% TITLE
%
% Enter your TITLE here
%%%%%%%%%%%%%%%%%%%%%%%%%%%%%%%%%%%%%%%%%%%%%%%%%%%%%%%%%%%%%%%%%%%%%
\title{\textbf{\large{Changes in zonal surface temperature gradients and Walker circulations in a wide range of climates}}}
%
% Author names, with corresponding author information. 
% [Update and move the \thanks{...} block as appropriate.]
%
\author{\textsc{Timothy M. Merlis}
				\thanks{\textit{Corresponding author address:} 
				Timothy M. Merlis, California Institute of Technology,
				1200 California Blvd., MC 100-23, Pasadena, CA 91125.
				\newline{E-mail: tmerlis@caltech.edu}}\quad\textsc{and Tapio Schneider}\\
\textit{\footnotesize{California Institute of Technology, Pasadena, California}}}

%
% Formatting done here...Authors should skip over this.  See above for abstract.
\ifthenelse{\boolean{dc}}
{
\twocolumn[
\begin{@twocolumnfalse}
\amstitle

% Start Abstract (Enter your Abstract above.  Do not enter any text here)
\begin{center}
\begin{minipage}{13.0cm}
\begin{abstract}
	\myabstract
	\newline
	\begin{center}
		\rule{38mm}{0.2mm}
	\end{center}
\end{abstract}
\end{minipage}
\end{center}
\end{@twocolumnfalse}
]
}
{
\amstitle
\begin{abstract}
\myabstract
\end{abstract}
}
%%%%%%%%%%%%%%%%%%%%%%%%%%%%%%%%%%%%%%%%%%%%%%%%%%%%%%%%%%%%%%%%%%%%%
% MAIN BODY OF PAPER
%%%%%%%%%%%%%%%%%%%%%%%%%%%%%%%%%%%%%%%%%%%%%%%%%%%%%%%%%%%%%%%%%%%%%
\section{Introduction}

The evolution of zonal surface temperature gradients in the tropics
is of interest in the study of future and past climates 
climates \citep[e.g.,][]{knutson95, fedorov06}. 
Recently, the changes in the
zonally asymmetric component of the tropical overturning circulation,
the Walker circulation, have been highlighted as a robust response
to warming in climate change simulations with comprehensive GCMs
\citep{held06, vecchi07b}.
Here, we examine the response of these zonally asymmetric aspects 
of the mean tropical climate to radiative perturbations in an idealized 
GCM and test theoretical arguments designed to
account for their changes.

How zonal surface temperature gradients evolve as the climate warms is 
unclear. On one hand, increased upwelling of cold
water on the eastern ocean margins may lead to increased zonal temperature
gradients in transient warmings \citep{clement96}, and this mechanism appears
to be important in coupled GCM simulations \citep{held10}.
On the other hand, paleoclimate evidence from the Pliocene
%taken at face value \citep[see][for an alternative interpretation]{wunsch09},
suggests substantially weakened or collapsed zonal temperature gradients in
a climate
only somewhat warmer than Earth's current climate \citep[e.g.,][]{fedorov06}.
(However, see \citet{wunsch09} for an alternative interpretation
of the data.)
%\citep[However, see][for an alternative interpretation of the data.]{wunsch09}
The multi-model mean of the fourth Intergovernmental Panel
on Climate Change (IPCC) assessment models shows little change in zonal 
temperature gradients in the Pacific basin over the twenty-first 
century \citep{dinezio09}.
As a step towards understanding the full climate dynamics that control
the zonal surface temperature gradient in the tropics, 
we develop a scaling relationship
based on the surface energy budget %that assumes knowledge of the 
%zonally asymmetric component of the divergence of the ocean energy flux 
and compare it to the results of the idealized GCM simulations.

The connection between changes in the hydrological cycle and 
large-scale atmospheric circulations in response to radiative 
forcing, for instance, by changed greenhouse gas 
concentrations, can be subtle. For example, 
the mass fluxes in the Hadley cell and extratropical eddy kinetic 
energy can vary non-monotonically as a longwave absorber is systematically
varied in an idealized GCM, with implications
for water vapor fluxes and the distribution of evaporation and precipitation
\citep{ogorman08b,ogorman08c,schneider10a,levine11a}.

Analysis of comprehensive GCM simulations suggests the
connection between changes in the hydrological cycle and 
upward mass fluxes is straightforward \citep{held06,vecchi07b}.
The argument is that precipitation, 
which is constrained by the energy balance at the surface
in the global mean, 
increases with warming more slowly than near-surface specific humidity,
which increases similarly to the rate given by the Clausius-Clapeyron 
relation for small changes in relative humidity; thus, the upward
vertical velocity that gives rise to condensation of water vapor weakens.
This scaling argument is sometimes interpreted to account for changes in the
Walker circulation \citep[e.g.,][]{vecchi06}, in part because 
changes in the zonally asymmetric component of the circulation
are larger than those in the zonally symmetric component \citep{held06}.
A difficulty in applying the hydrological cycle argument directly to 
local circulations, such as the Walker circulation, is that the 
surface energy balance constraint on the precipitation is lost, 
as water vapor convergence depends locally on the atmospheric circulation.
A central goal of this work is to assess the extent to which
the hydrological cycle arguments apply locally to the Walker circulation, 
and to compare this to global-mean versions of the 
scaling arguments to determine the extent to which they are adequate.

Here, we use an idealized GCM to systematically examine the response
of the zonally asymmetric tropical surface temperature and the 
Walker circulation to changes in the atmosphere longwave opacity. 
A scaling estimate based on the surface energy balance accounts for the
changes in the zonal surface temperature gradient. 
We assess the hydrological cycle estimate for changes in the Walker
circulation by applying it locally. The scaling estimate is accurate, though
it requires knowledge of the precipitation field.
In contrast, the global-mean version of the scaling estimate does not
in general account for the Walker circulation changes.

\section{Idealized GCM}

\begin{figure*}[t]
\begin{center}
  \noindent\includegraphics[width=33pc,angle=0]{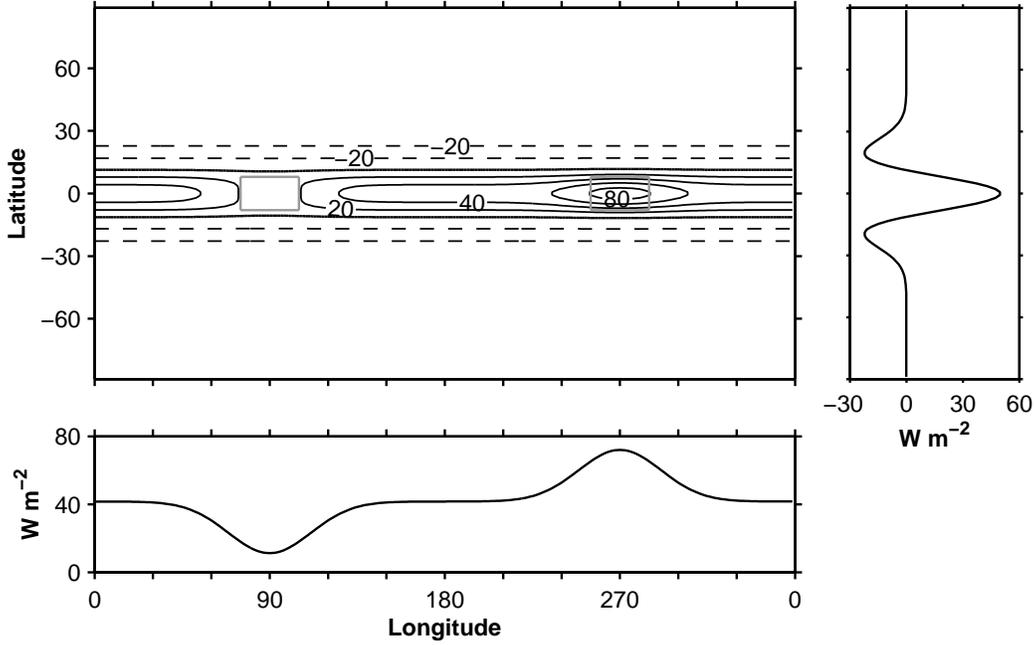}\\
\end{center}
  \caption{
    Ocean energy flux divergence, Q-flux, with positive values 
    indicate cooling tendencies. Large panel: longitude-latitude Q-flux 
    contours with contour interval of $20 \, \mathrm{W\,m^{-2}}$ and 
    negative values are dashed. The gray rectangles indicate the 
    averaging regions.
    Right panel: latitude vs. zonal-mean Q-flux. 
    Bottom panel: equatorial Q-flux (averaged within $8^\circ$ of the equator)
    vs. longitude.
  }
\label{fig-qflux}
\end{figure*}

\subsection{Model description}

We use an idealized, moist primitive-equation atmospheric GCM
similar to those described in \citet{ogorman08b} and \citet{frierson06a}.
Briefly, the model includes an active hydrological cycle with a 
simplified Betts-Miller convection scheme \citep{frierson07b}.
The scheme relaxes over a fixed timescale convectively unstable 
columns to moist pseudoadiabatic temperatures with 
constant ($70\%$) relative humidity. It has a 
gray longwave radiation scheme, and a slab-ocean surface boundary condition
with the heat capacity of $1 \, \mathrm{m}$ of water. 
The top-of-atmosphere insolation is an idealized, annual-mean profile,
and there is shortwave absorption in the atmosphere, both of which 
are as described in \citet{ogorman08b}. There is no 
condensed water phase in the atmosphere (i.e., there are no clouds). 
The climate is perturbed by varying the optical depth in the
gray radiation scheme by a multiplicative factor
to mimic variations in the trapping of longwave radiation due to 
variations in greenhouse gas concentrations.
The primitive equations are integrated using the spectral transform method
in the horizontal with T85 resolution (but the simulation results
are similar with T42 resolution). Finite differences in $\sigma = p/p_s$ 
(with pressure $p$ and surface pressure $p_s$) discretize the vertical
coordinate, with $30$ unevenly spaced levels. The time-stepping is 
semi-implicit, with a leap frog timestep of $300 \, \mathrm{s}$.
The simulation statistics are averaged over the last $1100 \, \mathrm{days}$
of the simulation following a spinup period of either $700 \, \mathrm{days}$ 
for an isothermal, resting initial condition or $300 \, \mathrm{days}$ for
an initial condition from an equilibrated simulation with similar optical 
thickness.
Aside from the horizontal resolution and timestep, the only difference 
between the simulations presented here and those presented in 
\citet{ogorman08b} is that we include a representation of ocean energy 
flux divergence in the slab ocean and present fewer simulated climates.

\subsection{Prescribed ocean energy flux divergence}

The surface boundary condition is a slab ocean that accounts 
for the radiative and turbulent surface fluxes. Additionally, it has 
a prescribed ocean energy flux divergence
that represents the influence of the ocean circulation on the 
surface temperature. This is commonly referred to as a Q-flux in
climate modelling.

The Q-flux consists of a component that varies only in latitude
and represents the divergence of the meridional ocean energy flux
\begin{equation}
\nabla \cdot F_0(\phi) =  Q_0 \ \left( \frac{1 - 2 \phi^2}{ \phi_0^2} \right) \ 
\mathrm{exp} \left(- \frac{ \phi^2 }{ \phi_0^2 } \right),
\end{equation}
where $\phi$ is latitude, $\phi_0 = 16^\circ$, and 
$Q_0 = 50 \, \mathrm{W m^{-2}}$ following \citet{bordoni07a} and 
\citet{bordoni08a}.
The meridional component of the Q-flux is 
necessary to have an Earth-like tropical circulation in the reference
climate; without it, the zonal-mean (Hadley) circulation is sufficiently 
strong that there is mean ascending motion everywhere along the equator
(i.e., the descending branch of the Walker circulation is weaker than
the Hadley circulation).

Zonal symmetry is broken by adding and subtracting Gaussian lobes along
the equator that are spaced $180^\circ$ longitude apart,
\begin{align}
 \nabla \cdot (F_0(\phi) + F_1(\lambda, \phi)) &= \nabla \cdot F_0(\phi) \notag \\
 &+ Q_1 \ \mathrm{exp} \left( -\frac{ (\lambda - \lambda_E)^2 }{ \lambda_1^2 }
 - \frac{ \phi^2}{\phi_1} \right)  \\
 &- Q_1 \ \mathrm{exp} \left( -\frac{ (\lambda - \lambda_W)^2 }{ \lambda_1^2 } 
 -  \frac{ \phi^2}{\phi_1} \right), \notag
\end{align}
where $\lambda$ is longitude, $\lambda_1 = 30^\circ$, $\lambda_E = 90^\circ$,
$\lambda_W = 270^\circ$, $\phi_1 = 7^\circ$, and 
$Q_1 = 40 \, \mathrm{W m^{-2}}$.

The resulting Q-flux, shown in Fig.~\ref{fig-qflux}, has similar magnitude
and spatial scale to reanalysis estimates 
\citep[e.g.,][their Fig.~5]{trenberth01b}.

\subsection{Series of simulations}

As in \citet{ogorman08b}, the longwave optical depth is varied 
to mimic changes in greenhouse gas concentrations. 
The optical depth $\tau = \alpha \tau_{\mathrm{ref}}$ is the product of a 
multiplicative factor $\alpha$
and a reference profile $\tau_{\mathrm{ref}}$ that depends on latitude and 
pressure,
\begin{equation}
\tau_{\mathrm{ref}} =[ f_l \sigma + (1 - f_l) \sigma^4 ] 
[\tau_{eq} + (\tau_p - \tau_{eq} ) \mathrm{sin}^2 \phi],
\end{equation}
where $f_l = 0.2$ and the reference optical
depths at the equator and pole are $\tau_{eq} = 7.2$ and $\tau_p = 1.8$,
respectively. We conduct simulations with rescaling factors $\alpha = 
(0.6$, $0.7$, $0.8$, $0.9$, $1.0$, $1.2$, $1.4$, $1.6$, $1.8$, $2.0$, $2.5$, 
$3.0$, $4.0$, $6.0)$.
The range of optical depths presented here is reduced compared to 
\citet{ogorman08b}; simulations with the smallest optical depths are 
omitted because the fixed meridional Q-flux leads to reversed Hadley cells and 
concomitant double intertropical convergence zones in sufficiently 
cold climates. 
In total, fourteen simulations are presented here. 

\section{Zonal surface temperature gradients}

\subsection{Simulation results}

\begin{figure}[!tb]
\begin{center}
  \noindent\includegraphics[width=19pc,angle=0]{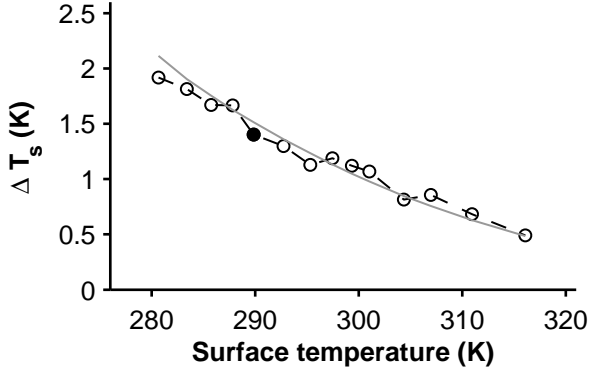}\\
\end{center}
\caption{
  Zonal surface temperature difference (west minus east in 
  Fig.~\ref{fig-qflux}) vs. global-mean surface temperature of GCM 
  simulations (circles with black dashed line), and scaling estimate
  (\ref{eqn-delta_temp_scaling}) (gray line). The averaging 
  conventions used to evaluate the scaling estimate are described in the text.
}
\label{fig-zonal_temp_diff}
\end{figure}

Figure~\ref{fig-zonal_temp_diff} shows the simulated zonal surface 
temperature differences averaged over the regions indicated in 
Fig.~\ref{fig-qflux} for each of the equilibrated climates, which are
identified by the time- and global-mean (denoted $\langle \cdot \rangle$)
surface temperature $\langle T_s \rangle$. Following \citet{ogorman08b},
the reference climate with $\alpha = 1.0$ is indicated by a filled symbol.
The east-west surface temperature difference $\Delta T_s$
decreases nearly monotonically as the climate warms and varies by 
more than a factor of three over the simulated range of climates.

The simulated zonal temperature difference in the reference climate 
($1.4 \, \mathrm{K}$) is smaller than that in Earth's current climate
($\about 5 \, \mathrm{K}$).
While we have used a Q-flux that has comparable magnitude to 
estimates for the current climate, the lack of clouds and the 
concomitant zonal asymmetry in cloud radiative forcing
leads to a smaller zonal asymmetry in the equilibrated surface temperature.
One remedy would be to impose a larger asymmetry in the Q-flux,
since in quasi-equilibrium theory, there is a near equivalence 
between surface fluxes and radiative forcing \citep[][section 2.3]{sobel10}.
However, we chose not to amplify the Q-flux asymmetry. 
Therefore, it is more meaningful to consider 
fractional changes in the simulated zonal surface temperature 
contrast than the magnitude of the simulated changes.
Our results hence may give changes in the surface temperature contrasts 
that, in reality, would likely be modified by processes we ignored, such 
as cloud-radiative feedbacks.

\subsection{Scaling estimate}

\begin{figure*}[!tb]
\begin{center}
  \noindent\includegraphics[width=33pc,angle=0]{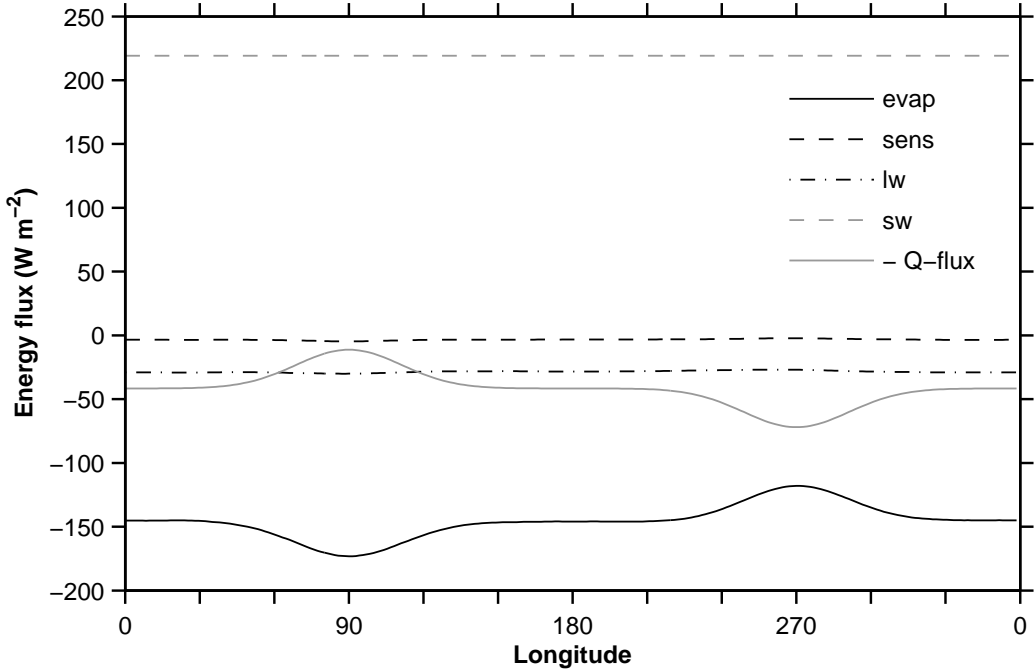}\\
\end{center}
\caption{
  Surface energy fluxes of the reference simulation vs. longitude 
  averaged within $8^\circ$ of equator. 
  ``lw'' and ``sw'' refer to the net longwave and shortwave
  radiative fluxes at the surface, respectively.
}
\label{fig-surf_energy_budget}
\end{figure*}

Figure~\ref{fig-surf_energy_budget} shows the surface energy budget as a 
function of longitude for the Earth-like reference simulation.
Examining the variations in longitude of the surface energy balance
reveals that the dominant balance in the zonally asymmetric component is 
between the prescribed Q-flux and the evaporative fluxes.
This is to be expected in Earth-like and warmer climates where the 
Bowen ratio (ratio of sensible to latent surface fluxes) is small
in the tropics. An analogous ratio can be formed for the net longwave
radiation at the surface \citep{pierrehumbert10}, and this is also small
for sufficiently warm climates.

This two-term balance between the zonally asymmetric
component of the Q-flux $\nabla \cdot F_1$ and evaporation 
holds over the range of climates we simulate, so that
\begin{equation}
\Delta \ ( \nabla \cdot F_1 ) \sim \Delta LE,
\label{eqn-two_term_balance}
\end{equation}
where $\Delta$ refers to east-west differences.

As the Q-flux (ocean energy flux divergence) is fixed, the bulk aerodynamic 
formula for evaporative fluxes can be manipulated with suitable assumptions 
to invert the zonal difference in the evaporative fluxes into a zonal difference
of the surface temperature.
The evaporative fluxes are computed by the bulk aerodynamic formula
\begin{equation}
E =  \rho c_d || {\bf v} || (q_s(T_s) - q) 
\approx \rho c_d || {\bf v} || (1 - \mathcal{H}) q_s(T_s),
\end{equation}
with %$L$ the latent heat of vaporization, 
air density $\rho$, drag coefficient $c_d$, surface wind speed $|| {\bf v} ||$,
saturation specific humidity $q_s$, specific humidity
of surface air $q$, surface temperature $T_s$, 
and surface relative humidity $\mathcal{H}$.
The second formula is approximate in that the ratio of the specific humidity
and the saturation specific humidity is set equal to the relative humidity
and in that the air-sea temperature difference is neglected. 
Neglecting the air-sea temperature difference is justifiable because
the subsaturation of air dominates over the air-sea temperature difference 
for conditions typical of tropical oceans.

For the climate changes we are examining, the east-west
evaporation difference then scales like the east-west difference
in the surface saturation specific humidity\footnote{
We are not suggesting that this scaling holds in general
(see, e.g., \citet{ogorman08b}
and our Fig.~\ref{fig-omega_scaling_decomp}
for a counterexample in the global mean),
but only for the zonal difference in the tropics.}
\begin{equation}
\Delta E \, \about \, \Delta q_s(T_s).
\label{eqn-deltaE_deltaqs}
\end{equation}
This is the result of the strong dependence of
the saturation specific humidity $q_s$ 
on temperature (fractional changes of $\about 7\% \, \mathrm{K^{-1}}$ 
for Earth-like temperatures). In contrast, changes in the surface relative humidity
$\mathcal{H}$ are relatively small. For example, \citet{schneider10a}
presented an argument based on a simplified surface energy budget that suggests
$\about 1\% \, \mathrm{K^{-1}}$ changes in $\mathcal{H}$. 
Regional changes that affect the zonal gradient of relative humidity
are potentially important, but neither comprehensive GCM simulations 
\citep{ogorman10} nor the idealized GCM simulations presented 
here exhibit substantial changes in the zonal relative humidity
gradient.
The neglected changes in the air-sea temperature difference and surface 
wind speed are also generally smaller than the changes in the surface 
saturation specific humidity.

To obtain an east-west surface temperature difference, rather than a saturation
specific humidity difference, we linearize the saturation specific
humidity
\begin{equation}
\Delta q_s(T_s) \approx \frac{\partial q_s}{\partial T} \Delta T_s.
\end{equation}
The resulting scaling estimate for the east-west temperature difference 
is the product of the east-west saturation specific humidity difference
$\Delta q_s$ and the inverse of the rate of change of the 
saturation specific humidity with temperature: 
\begin{equation}
\Delta T_s \sim \Delta q_s
\left( \frac{\partial q_s}{\partial T} \right)^{-1}.
\label{eqn-delta_temp_scaling}
\end{equation}
For invariant ocean energy fluxes, as in the idealized GCM simulations, 
$\Delta q_s$ is approximately constant by Eqns. (\ref{eqn-two_term_balance}) 
and (\ref{eqn-deltaE_deltaqs}), and 
the east-west surface temperature difference will decrease with warming 
by an amount given by the rate of change of the saturation specific humidity
with temperature (i.e., given by the Clausius-Clapeyron relation
as the surface pressure is nearly constant), evaluated using
a temperature that is representative of the surface in the tropics.

The decrease in the zonal temperature difference in a warming
scenario can be anticipated by considering the counterexample of 
invariant zonal surface temperature gradients: If the east-west 
temperature difference remained the same as the mean climate warmed, 
in the absence of changes in near-surface relative humidity, the east-west 
difference in evaporative fluxes would rapidly increase with warming. 
The resulting zonal asymmetry in the evaporation would be too large to be
balanced by an invariant ocean energy flux, and the other terms in the
surface energy balance are too small to restore equilibrium.
In a transient adjustment, the excess in evaporation would lead to a reduction
in the east-west surface temperature asymmetry as, for example, the large 
evaporation over warm regions would cool the surface.
The sensitivity of the evaporative fluxes to climate change and concomitant
reduction in zonal surface temperature gradients with warming was documented 
in the GCM simulations of \citet{knutson95}, who called it 
``evaporative damping''.

These approximations are adequate for the longwave radiation-induced
climate changes in the idealized GCM simulations, 
though they may not be in general. For example, 
in analyses of climate change simulations,
\citet{xie10} showed that changes in surface winds must be accounted 
for in the surface energy budget of some regions,
and \citet{dinezio09} showed that cloud-radiative feedbacks affect the
net surface shortwave radiation.

\subsection{Assessment of scaling estimate}

Fig.~\ref{fig-zonal_temp_diff} shows the scaling estimate
(\ref{eqn-delta_temp_scaling}) in gray.
The derivative of the saturation specific humidity with respect 
to temperature in (\ref{eqn-delta_temp_scaling})
is evaluated with the same simplified formulation for the 
saturation specific humidity as is used in the GCM, using the
simulated time- and zonal-mean surface temperature averaged within 
$8^{\circ}$ of latitude of the equator.
The saturation specific humidity difference $\Delta q_s$, evaluated 
over the averaging regions in Fig.~\ref{fig-qflux},
varies by %$\about 15\text{--}20\%$ over the range of simulations,
$\about 15\%$ over the range of simulations,
and we use the mean value, $2.4 \times 10^{-3}$, in 
the scaling estimate. %, eqn.~\ref{eqn-delta_temp_scaling}.
The magnitude of the variations in $\Delta q_s$ is consistent
with the magnitude of the neglected terms in the bulk aerodynamic
formula.
Overall, there is close agreement (maximum deviations $\about 15\%$) 
between the scaling estimate and the results of the GCM simulations.

We note that while the simulations and scaling estimate 
feature rapidly decreasing zonal temperature gradients as the
mean temperature increases ($7.4\% \, \mathrm{K^{-1}}$ at the 
reference climate), the decreases are not as rapid as proxies of
the Pliocene suggest ($\about 15\text{--}30\% \, \mathrm{K^{-1}}$). 
But it is possible that the neglected radiative and ocean feedbacks 
increase the sensitivity to changes.

\section{Walker circulation} \label{sec-walker}

\subsection{Simulation results}

\begin{figure}[!tb]
\begin{center}
  \noindent\includegraphics[width=19pc,angle=0]{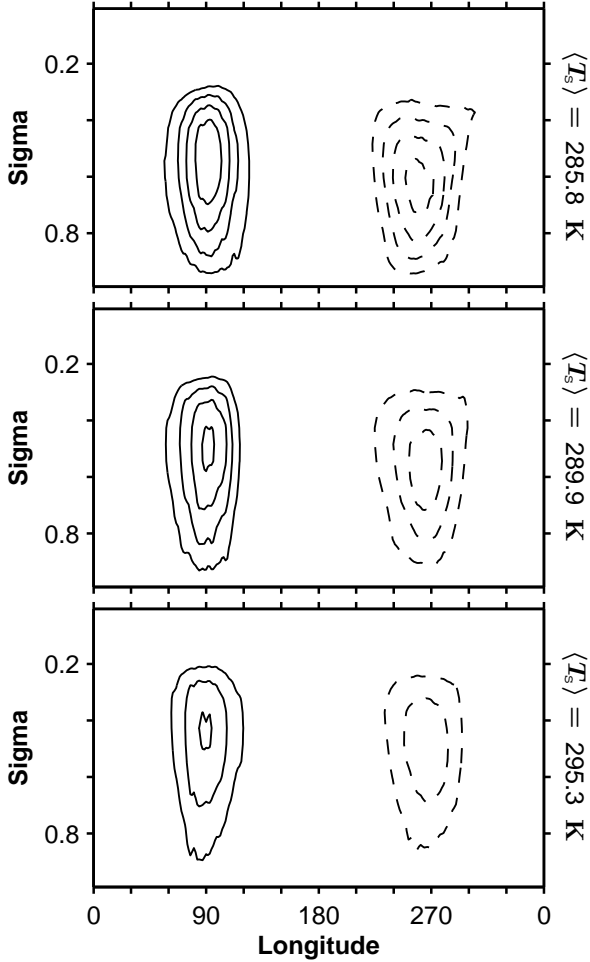}\\
\end{center}
\caption{
 Contours of the zonally asymmetric component of the time-mean 
 pressure velocity $\overline{\omega^*}$ in the longitude-sigma plane
 averaged within $8^\circ$ of the equator for three simulations
 with global-mean surface temperature $\langle T_s \rangle$ 
 indicated on the right. The contour interval is
 $0.01 \, \mathrm{Pa \, s^{-1}}$ with positive values (descending motion) 
 dashed. The zero contour is omitted.  
}
\label{fig-three_omega}
\end{figure}

Figure~\ref{fig-three_omega} shows the zonally asymmetric component of 
the time-mean pressure velocity, $Dp/Dt = \omega$, in the longitude-sigma 
plane for three simulations\footnote{All figures show $p_s D\sigma/Dt$, which
differs only slightly ($\sim 0.1\%$ in the region of interest) from $Dp/Dt$
as $\sigma Dp_s/Dt$ is small.}.
The zonally asymmetric pressure velocity is large over the regions of 
perturbed Q-flux; it is small elsewhere. The pressure depth over which
it has large amplitude increases with warming because the
tropopause height increases; its magnitude decreases with warming.
Note that a consequence of the Walker circulation's localization over
the Q-flux perturbations is that the averaging convention we have chosen
is adequate to capture nearly all of the ascending and descending 
mass fluxes---it is a closed circulation and alternative measures,
such as surface pressure differences, yield similar results.

\begin{figure*}[!tb]
\begin{center}
  \noindent\includegraphics[width=33pc,angle=0]{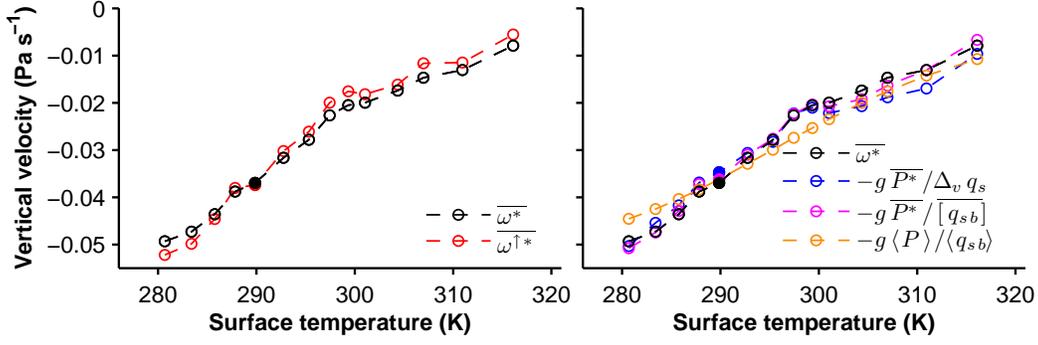}\\
\end{center}
\caption{
 Left: zonally asymmetric component of the pressure velocity 
 $\overline{\omega^*}$ (black) and zonally asymmetric component of the
 upward pressure velocity $\overline{\omega^{\uparrow *}}$
 (red) averaged over the ascending region of the Walker circulation
 (cf. Fig.~\ref{fig-qflux}). 
 The pressure velocity $\overline{\omega^*}$ is evaluated on the sigma level 
 where it is minimum (strongest ascent), and $\overline{\omega^{\uparrow *}}$ 
 is evaluated  on the same level. The upward velocity has been 
 multiplied by a factor of $1.6$ determined by minimizing the mean squared 
 deviation between $\overline{\omega^{\uparrow *}}$ and $\overline{\omega^*}$.
 Right: zonally asymmetric component of the pressure velocity, 
 $\overline{\omega^*}$, (black) and hydrological cycle scaling estimates:
 $-g \, \overline{P^*}/\Delta_v q_s$ (blue), 
 $-g \, \overline{P^*}/ \overline{ [q_{sb}] }$ (magenta), and 
 $-g \, \langle P \rangle / \langle q_{sb} \rangle$ (orange).
 The scaling estimates are multiplied by constants of 
 $1.5, 1.7$, and $0.8$, respectively, chosen to minimize the mean squared
 deviations between the scaling estimates and $\overline{\omega^*}$.
 The near-surface saturation specific humidity is evaluated on a 
 fixed sigma level $\sigma = 0.96$.
 The free-tropospheric contribution to the saturation specific 
 humidity difference $\Delta_v q_s$ is evaluated on a sigma level that 
 is a fixed amount larger than the sigma level 
 of the tropopause, $\sigma_t$: $\sigma = \sigma_{t} + 0.15$.
 The scale estimates $\overline{P^*}/\Delta_v q_s$ and $\overline{P^*}/\overline{ [q_{sb}] }$
 are evaluated within $8^\circ$ of the equator and 
 $\langle P \rangle / \langle q_{sb} \rangle$ is evaluated using global means.
 Zonal averages are used for $\Delta_v q_s$ and $\overline{ [q_{sb}] }$.
}
\label{fig-omega_scaling}
\end{figure*}

Figure~\ref{fig-omega_scaling} shows that the Walker circulation
mass flux in the ascending region 
rapidly weakens with warming. The fractional rate of decrease is 
$4.4 \% \, \mathrm{K^{-1}}$ at the reference climate relative to the
global-mean surface temperature. (The fractional change is larger
relative to local surface temperature changes as the warming
is greater in high latitudes.) 
The Walker circulation varies by a factor of six over the 
simulated range of climates.
In this figure, the zonally asymmetric component of the pressure velocity
%$\overline{\omega^*}$ 
is evaluated on the sigma surface where it is maximum, which varies from
$\sigma = 0.60$ in the coldest climate to $\sigma = 0.32$ in the warmest
climate. These changes roughly follow those of the zonal-mean
tropical tropopause (averaged within $8^\circ$ of the equator)
which rises from $\sigma = 0.24$ to $\sigma = 0.03$ 
determined by the World Meteorological Organization (WMO) 
$2 \, \mathrm{K \, km^{-1}}$ lapse rate criterion.
If the Walker circulation is evaluated using fixed sigma levels
in the lower troposphere, the weakening with warming is more rapid owing
to the combined changes in the vertical structure of the circulation and
the changes, shown in Fig.~\ref{fig-omega_scaling}, that are independent of 
the vertical structure.

\subsection{Scaling estimate}

In saturated ascending motion, the vertical advection of 
saturation specific humidity $q_s$ leads to condensation $c$,
\begin{equation}
- \omega^\uparrow \partial_p q_s \approx c,
\end{equation}
where $\omega^\uparrow = \omega H(-\omega)$ is a truncated upward pressure
velocity and $H$ is the Heaviside step function \citep{schneider10a}.
For a mass-weighted vertical average taken from the lifting condensation
level to the tropopause (denoted by $\{ \cdot \}$), the precipitation $P$ 
balances the vertical advection of specific humidity:
\begin{equation}
- \left\{ \omega^\uparrow \partial_p q_s \right\} \approx P.
\label{eqn-wv_adv}
\end{equation}
Equation~\ref{eqn-wv_adv} assumes that we are averaging 
over horizontal scales that are large enough to include convective 
downdrafts and the associated evaporation of condensate.
Neglecting transients aside from those implicitly included by using
the truncated vertical velocity, equation (\ref{eqn-wv_adv}) can be 
decomposed into zonal-means, denoted by $[ \cdot ]$, and deviations 
thereof, denoted by $( \cdot )^*$. The time-mean, denoted by
$\overline{ ( \cdot )}$, of the zonally asymmetric component of the budget is
then
\begin{equation}
  - \left\{ \overline{\omega^{\uparrow *}} \partial_p \overline{ [q_s] }  + 
  \overline{\left[ \omega^\uparrow \right] } \partial_p \overline{q_s^*} + 
  \overline{\omega^{\uparrow *}} \partial_p \overline{q_s^*} -
  \left[ \overline{\omega^{\uparrow *}} \partial_p \overline{q_s^*} \right]
  \right\} \approx \overline{P^*}.
\end{equation}
Terms 2-4 on the left-hand side, which include zonal asymmetries in 
saturation specific humidity $\overline{q_s^*}$,
are negligible as a result of the weak temperature gradients in the 
free troposphere of the tropics \citep{charney63,sobel01}. 
In the simulations, these terms combined are an order of magnitude smaller 
than the terms in the dominant balance

\begin{equation}
- \left\{ \overline{\omega^{\uparrow *}} \partial_p \overline{ \left[ q_s \right]} \right\}
\approx \overline{P^*}.
\label{eqn-asym_wv_dominant_bal}
\end{equation}

Our interest is in the zonally asymmetric component of the tropical
overturning circulation (denoted $\overline{\omega^*}$), the Walker 
circulation, and so we estimate the mass fluxes of the ascending or 
descending branches from (\ref{eqn-asym_wv_dominant_bal}) as
\begin{equation}
- \frac{ \overline{\omega^*} }{g} \about \,
- \frac{ \overline{\omega^{\uparrow *}} }{g} \about \,
\frac{ \overline{P^*} }{\Delta_v q_s},
\label{eqn-local_hydro}
\end{equation}
where $\Delta_v q_s$ is a zonal-mean vertical saturation specific humidity 
difference. 

This is closely related to the thermodynamic budget in the tropics
where the stratification is close to moist adiabatic.
On a moist adiabat (at constant saturation equivalent
potential temperature $\theta^*_e$), the vertical gradients of 
potential temperature $\theta$ and saturation specific humidity (assuming
constant latent heat of vaporization) are 
related by
\begin{equation}
-(T/\theta) \partial_p \theta |_{\theta^*_e} \approx (L/c_p) \partial_p
q_s |_{\theta^*_e}
\label{eqn-pottemp_shum_ma}
\end{equation}
\citep[e.g.,][]{iribarne81}.
Thus, as discussed in \citet{held06} and \citet{schneider10a}, the water 
vapor budget and thermodynamic budget are closely linked. The 
zonally asymmetric component
of both budgets is an approximate two-term balance between 
vertical advection of the zonal-mean potential temperature or 
saturation specific humidity by the zonally asymmetric velocity and the 
zonally asymmetric precipitation. 

\citet{held06} suggested that the free-tropospheric saturation specific
humidity in scaling estimates for the upward mass flux is negligible
or, as in \citet{betts89} and \citet{betts98}, it is linearly related to the 
boundary layer saturation specific humidity $\overline{ [q_{sb}] }$. 
In equation (\ref{eqn-local_hydro}), 
$\Delta_v q_s \about \, \overline{ [q_{sb}] }$ leads to the scaling estimate
\begin{equation}
- \frac{ \overline{\omega^*} }{g} \about \,
- \frac{ \overline{\omega^{\uparrow *}} }{g} \about \,
\frac{ \overline{P^*} }{\overline{ [q_{sb}]}},
\label{eqn-held_truncation}
\end{equation}

\citet{schneider10a} argued that the free-tropospheric contribution 
is not generally negligible---that is, an actual difference in saturation
specific humidity must be considered in (\ref{eqn-local_hydro}).
Physically, this can occur as the result of the amplification of 
warming aloft for moist adiabatic stratification. 
However, the amplification of temperature increases at fixed pressure levels
can be offset by increases in the pressure depth of the
mass fluxes, as occur with a rising tropopause. A concrete example 
of offsetting changes in stratification and pressure depth
is given by the ``Fixed Anvil Temperature'' hypothesis \citep{hartmann02},
which posits that the temperature that convection reaches is 
approximately climate invariant. If this
hypothesis holds precisely, the free-tropospheric saturation specific
humidity at the tropical tropopause would also be climate invariant.
\citet{schneider10a} also showed that a scaling that includes the
free-tropospheric contribution to the saturation specific humidity
difference accounted better for changes in the upward component of 
zonal-mean tropical circulation
in simulations with zonally symmetric forcing and boundary conditions.

Whether or not the free-tropospheric water vapor contributes depends on
the vertical velocity profile $\Omega(p)$ that is used to separate 
the integrand: $\{ \omega^\uparrow \partial_p q_s \} \approx 
\hat{\omega}^\uparrow \{ \Omega(p) \partial_p q_s \}$, 
where $\hat{\omega}^\uparrow$ depends on the horizontal, but not 
the vertical, coordinates \citep[e.g.,][]{sobel07a}.
This is discussed further in the appendix.

\subsection{Assessment of scaling estimate} \label{sec-wv_assess}

The left panel of Fig.~\ref{fig-omega_scaling} shows that the
zonally asymmetric component of the pressure velocity in the ascent
region scales with
the zonally asymmetric component of the upward pressure velocity:
$\overline{\omega^*} \sim \overline{\omega^{\uparrow *}}$.
However, the zonally asymmetric total pressure velocity $\overline{\omega^*}$ 
is systematically larger (by about $60\%$) than 
$\overline{\omega^{\uparrow *}}$.
The zonal-mean upward vertical velocity $\overline{[\omega^{\uparrow}]}$
is larger than the zonal-mean total vertical velocity $\overline{[\omega]}$;
hence, when the zonal-mean is removed to form the zonally asymmetric component 
(e.g., $\overline{\omega^*} = \overline{\omega} - \overline{[\omega]}$), 
the total velocity $\overline{\omega^*}$ is larger than the upward 
velocity $\overline{\omega^{\uparrow *}}$.
The fact that $\overline{\omega^*} \sim \overline{\omega^{\uparrow *}}$ suggests that 
if the hydrological cycle scaling estimate accounts for the changes in the 
upward pressure velocity, to which it is more directly related 
(\ref{eqn-wv_adv}), it will also account for the changes in the 
total pressure velocity, i.e., the Walker circulation.
However, consistent with the results of \citet{schneider10a},
the simulated changes in the zonal-mean pressure velocity
$\overline{[\omega]}$ in the ascending branch of the Hadley
circulation are not straightforwardly related to the changes in the upward 
component of the zonal-mean pressure velocity 
$\overline{[\omega^\uparrow]}$.

\begin{figure*}[!tb]
\begin{center}
  \noindent\includegraphics[width=33pc,angle=0]{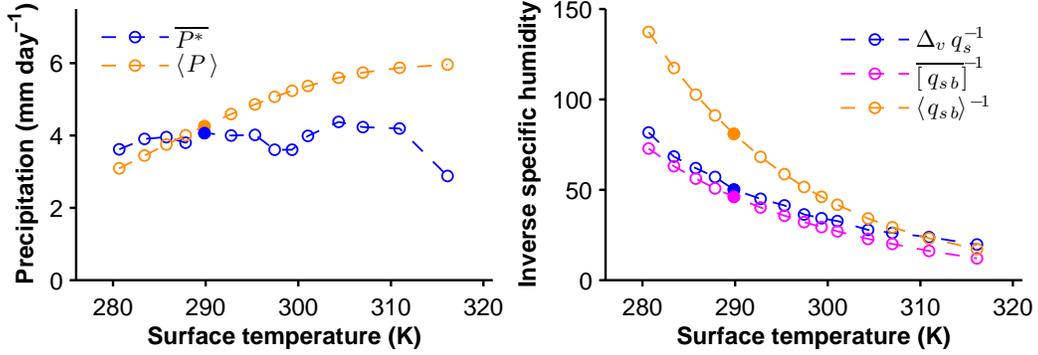}\\
\end{center}
\caption{
 Left: precipitation vs. global-mean surface temperature for 
 zonally asymmetric tropical-mean $\overline{P^*}$ (blue) and 
 global-mean $\langle P \rangle$ (orange).
 Right: inverse specific humidity vs. global-mean surface temperature for
 tropical-mean $\Delta_v q_s$ (blue), tropical-mean $\overline{ [q_{sb}] }$ (magenta), and 
 global-mean $\langle q_{sb} \rangle$ (orange).
 The averaging conventions are the same as in Fig.~\ref{fig-omega_scaling}.
}
\label{fig-omega_scaling_decomp}
\end{figure*}

The right panel of Fig.~\ref{fig-omega_scaling} shows that the 
hydrological cycle
estimate (\ref{eqn-local_hydro}) applied locally accounts for the
changes in the Walker circulation.
There are two variants of the local scaling in the right panel of 
Fig.~\ref{fig-omega_scaling}: one includes the free-tropospheric
saturation specific humidity contribution to $\Delta_v q_s$ (blue)
and one that neglects it 
$\Delta_v q_s \about \, \overline{ [q_{sb}] }$ (magenta).
The two are largely indistinguishable and both account for the simulated
changes in $\overline{\omega^*}$. 
The scaling that neglects the free-tropospheric saturation specific
humidity is an approximation, so while it appears to be
more accurate in the warm limit, this is the result of compensating errors.
The scaling constants, obtained by least squares, are approximately $1.5$
which is consistent with the factor of $1.6$ that relates
$\overline{ \omega^{\uparrow *} }$ and $\overline{ \omega^*}$. 

Figure~\ref{fig-omega_scaling} also shows the global-mean scaling
estimate, $- g \, \langle P \rangle / \langle q_{sb} \rangle$,
in orange. 
This is similar to the convention used in the analysis of IPCC simulations
in \citet{held06} and \citet{vecchi07b}, who further approximated 
the saturation specific humidity changes by linearizing its dependence
on temperature about the present climate (i.e., $\delta \langle q_{sb} \rangle
\approx 0.07 \times \delta \langle T_s \rangle$).
The global-mean scaling captures the gross magnitude of the weakening, 
but not the detail.
It tends to underestimate the circulation changes 
for climates colder than the reference climate;
at the reference climate, the global-mean scaling decreases by 
$3.1\% \, \mathrm{K}^{-1}$, compared with $\overline{\omega^*}$ which 
decreases by $4.4\% \, \mathrm{K}^{-1}$.

As the local scaling convention captures the variations in the Walker
circulation with climate, we can decompose it to determine the relative
contributions from the precipitation changes 
(Fig.~\ref{fig-omega_scaling_decomp}, left panel) and the saturation
specific humidity changes (Fig.~\ref{fig-omega_scaling_decomp}, right panel).
Two factors---approximately constant $\overline{P^*}$ and rapidly
decreasing $\Delta_v q_s^{-1}$ or $\overline{ [q_{sb}] }^{-1}$---account 
for the rapid decrease in the Walker circulation with warming.

The zonally asymmetric component of tropical precipitation, $\overline{P^*}$, 
is approximately constant with climate; however, this is not a general result.
Alternative convection scheme parameters, in particular relaxing to 
$90\%$ relative humidity instead of $70\%$, 
can produce non-monotonic changes in $\overline{P^*}$ with climate.
The locally averaged version of the scaling for Walker circulation
changes (\ref{eqn-local_hydro}) is similarly accurate in the 
alternate set of simulations.
These simulations have increases in $\overline{P^*}$ and 
a less rapid decrease in the Walker circulation $\overline{\omega^*}$
with warming from the Earth-like reference climate than the simulations 
presented here.
The multi-model mean of the models participating in the fourth IPCC
assessment also features increased $\overline{P^*}$ with warming
\citep{meehl07}.

Changes in local precipitation cannot be understood 
independently of the circulation changes as local water vapor convergence
depends, in part, on the circulation changes. 
In the simulations presented here,
the changes in the zonally-asymmetric component of net 
precipitation in the deep tropics have a large dynamic component,
i.e., a component associated with changed winds and fixed specific humidity.
The dynamic changes reduce the zonally-asymmetric precipitation due 
to weakening surface easterlies with warming and offset the 
thermodynamic component of the changes, i.e., the component associated with 
changes in specific humidity and fixed winds, which increase the 
the zonally-asymmetric precipitation with warming.
This is qualitatively consistent with the multi-model mean of the 
fourth IPCC assessment simulations presented
in Fig. 7 of \citet{held06}: the equatorial Pacific is a region 
in which the changes in net precipitation are quite different from
a thermodynamic estimate that assumes fixed circulation.

The saturation specific humidity is a rapidly increasing function of 
temperature, so, as expected, its inverse rapidly decreases with warming.
The difference between the inverse saturation specific
humidity when the free-tropospheric component of $\Delta_v q_s$ is neglected
and when it is retained is small. This is a consequence of the depth of 
the circulation---it reaches sufficiently low pressures and temperatures 
that $q_s$ is negligible in that region of the free troposphere. 
The appendix has a more extensive discussion of
the dependence of the scaling estimate on the vertical structure of the 
circulation and how the results of the simulations presented here relate to
previous work.

Figure~\ref{fig-omega_scaling_decomp} also sheds light on why 
the global-mean convention may, in some cases, appear adequate.
Global-mean precipitation $\langle P \rangle$ increases more rapidly 
than the zonally asymmetric component of tropical precipitation 
$\overline{ P^* }$; global-mean 
surface saturation specific humidity $\langle q_{sb} \rangle$ 
increases more rapidly than the tropical-mean surface saturation 
specific humidity difference $\Delta_v q_s$. As the 
scaling depends on the ratio, the more rapid changes in the individual
components largely cancel, but this must be regarded as fortuitous.

\section{Conclusions}

Zonal surface temperature gradients decrease rapidly as the climate warms
in the idealized GCM.
This can be understood by considering a simplified surface energy budget
that consists of a two-term balance between the zonally asymmetric
component of the Q-flux and evaporation. 
It leads to a scaling for the zonal difference in tropical
surface temperature that depends inversely on the rate of change of 
saturation specific humidity (or the Clausius-Clapeyron relation), 
evaluated using the 
mean surface temperature in the tropics if the Q-flux is fixed. 

The Walker circulation
rapidly weakens with warming in the idealized GCM. 
The changes can be accounted for using locally evaluated hydrological 
cycle scaling estimates. The changes in the scalings can be decomposed 
into i) equatorial precipitation increases being approximately zonally 
uniform with warming and ii) rapid, Clausius-Clapeyron increases 
in near-surface saturation specific humidity with warming.
The global-mean version of the hydrological cycle scale
estimate does not accurately account for the changes in the Walker
circulation in all climates, though it does capture gross changes due to 
compensating changes in precipitation and saturation specific humidity that
are larger than the locally averaged quantities.

That the hydrological cycle scaling estimate requires knowledge of 
local precipitation changes is a limitation. 
Ideally, one would want to understand circulation changes independently and
use that result in the water vapor budget to constrain local precipitation
changes. The moist static energy budget is one approach, though it has 
established difficulties \citep{sobel07a}.
Another approach is that of the Lindzen-Nigam model \citep{lindzen87b},
in which the surface temperature field is used to determine pressure
gradients, which in combination with a linear momentum equation determine
boundary layer convergence.
We have solved the Lindzen-Nigam equations using the GCM simulated
surface temperature fields as forcing and found that the changes
the Lindzen-Nigam model predicted in the boundary layer mass flux 
underestimated those simulated by the GCM by roughly a factor of two 
over the range of climates. 
Devising conceptual models of how tropical circulations change with
climate that adequately account for GCM simulations remains a 
challenge.

A limitation of this study is the prescribed and invariant Q-flux.
But understanding the atmosphere-only dynamics is a necessary step towards
to the more realistic case that includes an interactive ocean.

\begin{acknowledgment} 
We thank Simona Bordoni, Isaac Held, Yohai Kaspi, Xavier Levine, and 
Paul O'Gorman for helpful discussions and technical assistance. 
The comments of two anonymous reviewers helped clarify
the presentation of our work.
This work was supported by a National Defense Science and
Engineering Graduate Fellowship, a National Science Foundation
Graduate Research Fellowship, and a David and Lucile Packard
Fellowship. The GCM simulations were
performed on Caltech's Division of Geological and Planetary Sciences
Dell cluster. The program code for the simulations, based on the
Flexible Modeling System of the Geophysical Fluid Dynamics Laboratory,
as well as the simulation results themselves are available from the
authors upon request.
\end{acknowledgment}

% Use appendix}[A], {appendix}[B], etc. etc. in place of appendix if you have multiple appendixes.
\ifthenelse{\boolean{dc}}
{}
{\clearpage}
\begin{appendix}
\section*{\begin{center} Dependence of water vapor scaling on vertical structure \label{sec-appendix} \end{center}}

\begin{figure}[!tb]
\begin{center}
  \noindent\includegraphics[width=19pc,angle=0]{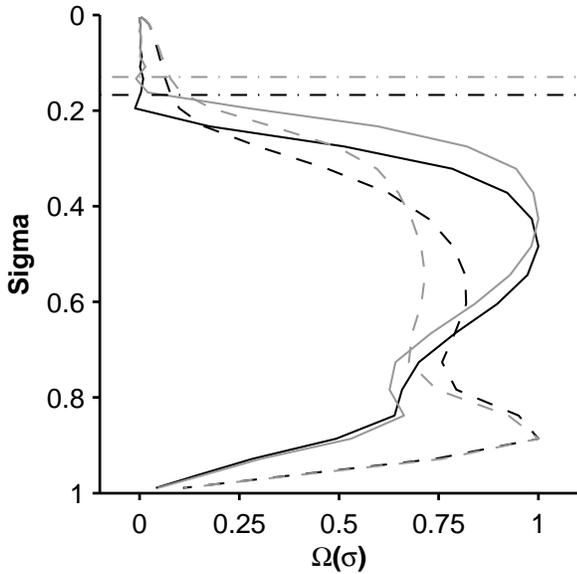}\\
\end{center}
\caption{
 Vertical structure of vertical velocity
 $ \Omega(\sigma) = \omega^\uparrow / \mathrm{min}(\omega^\uparrow) $
 for simulations with global-mean surface temperature
 $\langle T_s \rangle$ $289.9 \, \mathrm{K}$ (black) and 
 $295.3 \, \mathrm{K}$ (gray).
 Solid lines are zonally asymmetric pressure velocity profiles 
 $\overline{ \Omega^* }$ evaluated in the ascending region (cf. Fig.~\ref{fig-qflux})
 and dashed lines are zonal-mean pressure velocity profiles  $\overline{ [ \Omega ] }$.
 The dashed-dotted line is the zonal-mean tropical tropopause defined by the  
 WMO $2 \, \mathrm{K \, km^{-1}}$ lapse rate criterion.
}
\label{fig-omega_basis}
\end{figure}

To understand why the free-tropospheric contribution to the 
saturation specific humidity difference is negligible, we 
examine the vertical structure of the vertical velocity, as 
$\Delta_v q_s$ is an approximation of $\{ \Omega(p) \partial_p q_s \}$.
Also, the vertical structure of the circulation has importance beyond the 
hydrological cycle scaling estimate: it determines whether moist static energy 
is imported or exported by overturning circulations 
\citep{sobel07a, back06, peters08}.
If the vertical velocity profile has larger amplitude 
in the free troposphere, it makes the saturation specific humidity
difference less sensitive to the free-tropospheric values entering
$\Delta_v q_s$ because temperature and saturation specific humidities 
are small there. Indeed, the vertical velocity profile is ``top-heavy'' in 
the ascending branch of the Walker circulation (solid lines 
in Fig.~\ref{fig-omega_basis}); that is, it has significant 
amplitude in the upper troposphere. 
In contrast, the zonal-mean vertical velocity profile has
more amplitude near the surface and less amplitude 
in the upper troposphere (dashed lines in Fig.~\ref{fig-omega_basis}).
The vertical velocity profiles of the idealized simulations are 
broadly consistent with the regional variations in the vertical 
velocity profile in Earth's tropics \citep{back06}.

For the zonal-mean component of the simulations presented here,
the hydrological cycle scaling estimate that neglects the changes 
in the saturation specific humidity away from the boundary layer
($\Delta_v q_s \about \, \overline{ [q_{sb}] }$) is distinguishable from the
scaling estimate that retains the full $\Delta_v q_s$
(i.e., the representative pressure levels when the integral is discretized
are closer to each other than in the zonally asymmetric case 
and the free-tropospheric contribution is not negligible).
This is consistent with the results presented in \citet{schneider10a}.

Figure~\ref{fig-omega_basis} also shows that the pressure depth 
over which there is significant vertical velocity amplitude increases 
and the tropopause pressure decreases with warming.
For this reason, we vary the pressure at which the free-tropospheric 
contribution to $\Delta_v q_s$ is evaluated
in Fig.~\ref{fig-omega_scaling};
neglecting these changes can potentially lead to different scaling behavior.

There are two ways in which the vertical velocity profile can
lead to Clausius-Clapeyron-like scaling of $\Delta_v q_s$.
First, a larger pressure depth decreases the free troposphere's
weight (e.g., in the limit of the whole troposphere, the tropopause
saturation specific humidity is orders of magnitude smaller than that
of the surface). 
Second, holding the pressure depth fixed and shifting it upward (e.g., 
the difference between $800 \, \mathrm{hPa}$ and $300 \, \mathrm{hPa}$
compared to the difference between $900 \, \mathrm{hPa}$ and 
$400 \, \mathrm{hPa}$) leads to more rapid changes in the saturation 
specific humidity difference as the temperatures are lower and the rate 
of change of the saturation specific humidity is larger. These changes 
will appear closer to the rate given by the Clausius-Clapeyron relation 
if it is evaluated at a fixed level such as the surface 
\citep[related issues about the sensitivity of saturation
specific humidity to where it is evaluated and 
the possible imprecision that results are discussed in][]{ogorman10}. 

In an analysis of comprehensive GCMs, \citet{vecchi07b} showed 
that changes in the global-mean upward component of the vertical velocity
scaled with the hydrological cycle scaling estimate that neglects
the free-tropospheric contribution to the saturation
specific humidity (i.e., $\langle \omega^\uparrow \rangle \about \,
-g \, \langle P \rangle / \langle q_{sb} \rangle$), but
the scaling overestimated the magnitude of the vertical velocity changes
by about a factor of two to three. 
However, when \citet{vecchi07b} considered only upward pressure velocities
exceeding $0.05 \, \mathrm{Pa\,s^{-1}}$ 
(i.e., $\omega^\uparrow = \omega H( -0.05 \, \mathrm{Pa\,s^{-1}} - \omega)$),
the changes in the vertical velocity were larger and consistent with
the scaling estimate.
Changes in the vertical structure of the circulation
may explain why the scaling in Fig.~5 of \citet{vecchi07b} 
improved when a non-zero threshold for upward vertical velocity was 
introduced: With a threshold of zero, the basis function is less top-heavy and 
the changes in the free-tropospheric saturation specific humidity 
contribution to $\Delta_v q_s$ cannot be neglected, in which case
approximating $\Delta_v q_s$ with $\overline{ [q_{sb}] }$ leads to an overestimate
of the predicted weakening of the circulation with warming; however, 
for regions of strong ascent (such as those considered here), the 
vertical velocity profile can be sufficiently deep that the 
free-tropospheric contribution to $\Delta_v q_s$ is negligible,
$\Delta_v q_s$ scales similarly to $\overline{ [q_{sb}] }$, and the circulation weakens
rapidly with warming.
Another factor is that  
\citet{vecchi07b} changed from a threshold of $0 \, \mathrm{Pa\,s^{-1}}$,
which includes extratropical regions of ascending motion, to 
a threshold of $0.05 \, \mathrm{Pa\,s^{-1}}$, which limits the area 
considered to a few, strongly ascending tropical locations, while 
using global-mean precipitation and surface specific humidity in both cases. 
The utility of global-mean scaling estimates for local circulation 
changes is assessed in Fig.~\ref{fig-omega_scaling} and 
Fig.~\ref{fig-omega_scaling_decomp} which show that it does not generally account 
for the simulation results. 
Because changing the threshold changes both the geographic 
area under consideration and vertical structure,
it is unclear which is more important. 

\end{appendix}

% Create a bibliography directory and place your .bib file there.
\ifthenelse{\boolean{dc}}
{}
{\clearpage}
\bibliographystyle{./ametsoc}
\bibliography{./bibliography/references}
%\bibliography{./walker.bib}

%%%%%%%%%%%%%%%%%%%%%%%%%%%%%%%%%%%%%%%%%%%%%%%%%%%%%%%%%%%%%%%%%%%%%
% FIGURES
%%%%%%%%%%%%%%%%%%%%%%%%%%%%%%%%%%%%%%%%%%%%%%%%%%%%%%%%%%%%%%%%%%%%%
%\begin{figure}[t]
%  \noindent\includegraphics[width=19pc,angle=0]{./figs/qflux.eps}\\
%  \caption{Enter the caption for your figure here.  Repeat as
%  necessary for each of your figures. Create a figures directory and
%  place all figures in that directory. Figure from Houghton et al. (2001).}
%\label{fig-qflux}
%\end{figure}

\end{document}